\documentclass{PoS}
\usepackage{graphicx}
\usepackage{amsmath,amssymb,latexsym,wrapfig}
\usepackage{url}
\usepackage{bm}
\usepackage{xspace}
\usepackage{nicefrac}

\newcommand\ion[2]{#1$\;${\scshape{#2}}\relax}

\def\nii{[\ion{N}{ii}]}
\def\niilam{[\ion{N}{ii}]~$\lambda6584$}

\def\oiiilam{[\ion{O}{iii}]~$\lambda5007$}

\newcommand{\Hb}{\ensuremath{{\rm H}\beta}}
\newcommand{\Ha}{\ensuremath{{\rm H}\alpha}}

\newcommand{\beq}{\begin{equation}}
\newcommand{\eeq}{\end{equation}}

\newcommand{\be}{\begin{equation}}
\newcommand{\ee}{\end{equation}}

\newcommand{\apj}{"Astrophys. J."}

\title{A Uniformly Selected, All-Sky Optical AGN catalog, for UHECR Correlation}

\ShortTitle{Optical AGN Catalog}

\author{Ingyin Zaw$^{a,*}$, Yanping Chen$^{a,\dagger}$,  Glennys R. Farrar$^{b}$\\
 a) New York University, Abu-Dhabi \\
  b) Center for Cosmology and Particle Physics, New York University, NY, NY \\
  * E-mail: \email{ingyin.zaw@nyu.edu}  ~$\dagger$Speaker at ICRC 2015.}
        
\abstract{Studies discerning whether there is a significant correlation between UHECR arrival directions
and optical AGN are hampered by the lack of a uniformly selected and complete all-sky optical AGN
catalog. To remedy this, we are preparing such a catalog based on the 2MASS Redshift
Survey (2MRS), a spectroscopic sample of $\sim 44,500$ galaxies complete to a K magnitude of 11.75
over 91\% of the sky. We have analyzed the available optical spectra of these 2MRS galaxies
($\sim 80$\% of the galaxies), in order to identify the AGN amongst them with uniform criteria. We present a first-stage
release of the AGN catalog for the southern sky, based on spectra from the 6dF Galaxy survey and CTIO telescope. 
Providing a comparably uniform and complete catalog for the northern sky is more challenging because
the spectra for the northern galaxies were taken with different instruments.}

\FullConference{The 34th International Cosmic Ray Conference,\\
		30 July- 6 August, 2015\\
		The Hague, The Netherlands}



\begin{document}

\section{Brief sketch of contribution} 

We are building the first complete, uniformly selected catalog of optical active galactic nuclei (AGN).  
Optical AGN are a disjoint population from  X-ray AGN \cite{Arnold+XrayvsOpticalAGN09}, probably corresponding to different accretion states.  Thus from the standpoint of UHECR-acceleration, the two types of AGN should be considered distinct source candidates so that the optical AGN catalog provided here is complementary to the Swift-BAT X-ray AGN catalog for UHECR-AGN correlation studies. 

There is a tantalizing persistence of a correlation at the 2- to 2.7-sigma level between UHECRs and galaxies in the VCV catalog\footnote{Auger now reports approximately 2-sigma \cite{augerAniso14}, a similar significance as in their original report \cite{augerScience07}, and TA reports a 2.7-sigma level ({\url{http://uhecr2014.telescopearray.org/tinyakov/UHECR2014-anisotropy-tinyakov.pdf}}, updating \cite{TAaniso13}.}. However, the VCV catalog \cite{vcv} is a compendium of optical AGN candidates reported in the literature, with highly non-uniform sampling \cite{zfb10} so that in some regions of the sky many AGN are missing;  at the same time, the VCV catalog includes galaxies which have emission lines but are not AGN \cite{zfg09,tzf12}.   This makes it impossible to reliably evaluate the significance of reported correlations between VCV galaxies and UHECR arrival directions, and calls for creating a proper, complete and uniformly-selected catalog of optical AGN which is the motivation for the present work.

The Two Micron All-Sky Survey \cite{2MASS} 
identified all galaxies outside the Galactic Plane above a specified infra-red (2-micron) flux limit.  The 2MASS Redshift Survey (2MRS) \cite{2MRSz} contains redshifts of $\approx 44,500$ 2MASS galaxies brighter than K magnitude of 11.75 (mean $z = 0.03$), most obtained from optical spectra; thus 2MRS provides redshifts for a uniformly-selected catalog of galaxies, which is complete over 91\% of the sky. We use the 2MRS catalog as the parent sample for AGN identification. While 2MRS provides a uniformly selected and complete sample of galaxies, the spectra used for redshift determination are from several different sources \cite{2MRSz}, namely the Sloan Digital Sky Survey (SDSS), 6dF Galaxy Survey (6dF), ZCAT, and spectra taken specifically for 2MRS using the FLWO 1.5m telescope (FAST), the CTIO 1.5m telescope (CTIO), and the MacDonald Observatory. In addition, roughly a quarter of the redshifts were taken from the NASA Extragalactic Database (NED). This poses a challenge for constructing a uniform AGN catalog since the spectra from the different instruments have different signal-to-noise ratios and spectral resolution. For example, the SDSS spectra are of much higher quality than the rest. The NED sample in particular is problematic not only because the spectra come from hundreds of different sources but also because some of the redshifts are from non-optical sources, e.g. radio spectra of the 21 cm hydrogen line. 

We have obtained the digital optical spectra from SDSS, 6dF, FAST, CTIO, and a fraction of the NED sources, representing $\sim$80\% of the 2MRS spectra. Our first catalog consists of AGN in the southern sky for two reasons. First, the NED galaxies are mostly ($\sim$ 75\%) in the north. Second, and more importantly, most ($\sim$82\%) of the galaxies in the southern sky have spectra from 6dF or CTIO and the spectra from these two instruments are of similar quality. In the north, SDSS spectra have a much higher signal-to-noise than the spectra from other instruments. Consequently, the completeness of the AGN catalog will be more uniform in the south. Furthermore, VCV is more incomplete in the south than in the north since large AGN catalogs in the southern sky were unavailable before 6dF. Since optical light is absorbed by the material in our galaxy, we exclude the Galactic plane region (i.e. require $|b|  > 10^\circ$) from our sample. The galaxies with redshifts in the range 0.0407--0.0511 have contamination from telluric lines (due to the Earth's atmosphere) which fall within the \Ha\--\nii\ region and are, therefore, excluded from the sample. For UHECR correlation studies it is sufficient to restrict to $z < 0.0407$, since due to the GZK horizon very few observed UHECRs come from higher redshift than this, which corresponds to a co-moving radial distance of  173.8 Mpc for H$_{0} = 70$ km/s/Mpc.

Our method for AGN identification from optical spectra is described in Section 2.  We first subtract the stellar absorption lines and continuum emission of the host galaxy from the total spectrum. The resulting spectrum consists only of emission lines (if present). A galaxy is identified as an emission line galaxy if at least two of the lines used for AGN identification are present at the 90\% C.L. Broad line AGN (also known as Seyfert 1 or Type-1 AGN) exhibit characteristic broadening of the \Ha\ and \Hb\ lines due to the high-velocity orbital motion in the inner part of the accretion disk when the central region is seen face-on. If the \Ha\ broad component has a full width half maximum (FWHM) greater than 1000 km/s, we identify the galaxy as a Seyfert 1. If only narrow emission lines are present, we determine if the galaxy is a narrow line AGN, a.k.a. Seyfert 2 or Type 2, by comparing the ratios of \oiiilam\ to \Hb\ and \niilam\ to \Ha\ as first prescribed by Baldwin, Phillips, and Terlevich \cite{BPT81}, and is known as the BPT diagram. The plot of the \oiiilam/\Hb\ ratio against the \niilam/\Ha\ ratio is known as the BPT diagram. We identify as AGN, galaxies which satisfy either the more stringent criteria based on theoretical modeling (Kewley et al., 2001,\cite{Kewley01}) or the looser criteria from empirical fits to the SDSS sample (Kauffmann et al., 2003 \cite{Kauffmann03}), and indicate the basis of the classification.

The resulting southern sky catalog (dec $< 0^\circ$, $z < 0.0407$, and $|b| >10^\circ$) contains 351 broad line AGN, 1914 narrow line AGN satisfying the Kewley \cite{Kewley01} criteria, and 4238 narrow line AGN satisfying the Kauffmann \cite{Kauffmann03} criteria. The homogeneity and completeness of the catalog are discussed in Section 3. The length of the proceedings does not permit us to list the full catalog but it can be found at {\tt \footnotesize http://nyuad.nyu.edu/en/academics/faculty/ingyin-zaw/agn-catalog.html}. Our catalog contains 2.7 (5.5 with Kauffmann criteria) times more AGN than VCV  and is much more uniform both across the sky and in redshift, and is currently the most complete and uniform optical AGN catalog available for the southern sky. 



\section{AGN Identification}

The first step in the preparation of the catalog is to automate the subtraction of the host-galaxy contribution to the spectrum.  For each galaxy, the host galaxy spectrum is determined allowing emission line feature to be identified, by subtracting the best-fit superposition of spectra from stellar population models \cite{Chen13}, based on the first release of the X-shooter Spectral Library (XSL)\cite{Chen14} using the full-spectrum-fitting 
program pPXF \cite{Cappellari04}.  With the wide wavelength coverage (3000 - 25000 \AA) of the X-shooter library, all the absorption stellar features can be fit simultaneously.  We mask the emission line regions and any bad pixel regions for the full-spectrum-fitting.  Since XSL is an empirical library, there is no model dependence on line profiles or opacity uncertainty. We use a 3rd order polynomial to modulate the large-scale spectral shape, since the CTIO and 6dF data are not flux calibrated.

\begin{figure}[t]
\centering
\vspace*{-1.0cm}
\includegraphics[width=.8\textwidth]{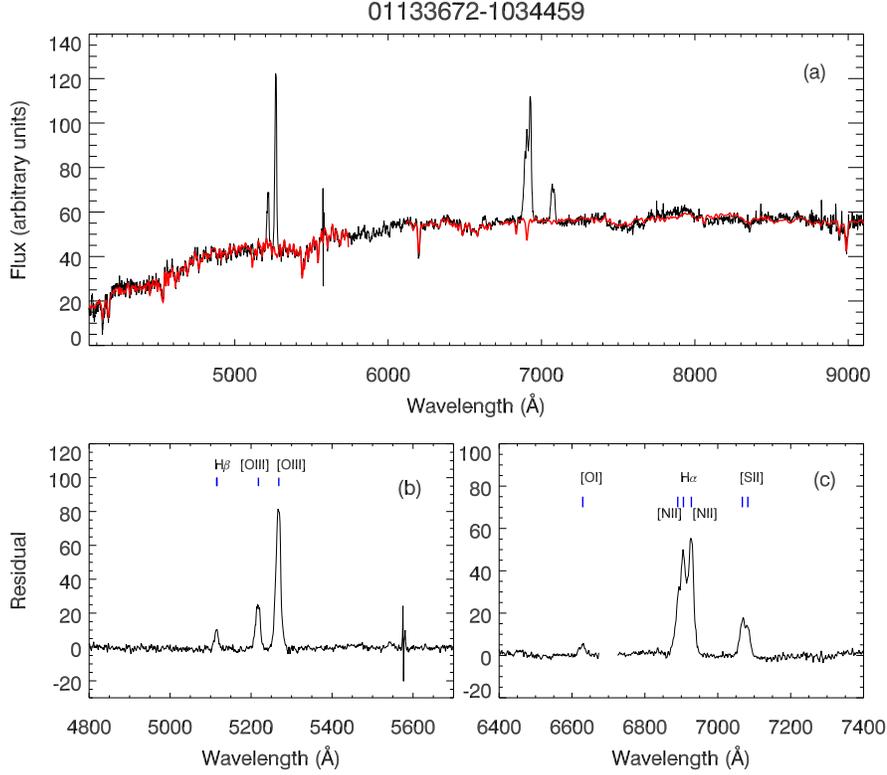}
\caption{Full-spectrum-fitting of an illustrative SDSS DR8 spectrum. Panel (a): DR8 data as black line, with the best host-galaxy spectrum model fit in red. 
The gap in the model spectrum is due to the non-recoverable dichroic features, see \cite{Chen14} for details.
Panel (b) and (c): The residual spectrum zoomed in on two wavelength ranges having the interesting emission features.}
\label{dr8ssp}
\vspace*{-0.5cm}
\end{figure}

Figure~\ref{dr8ssp} shows an example of a spectrum (black) from SDSS data release 8 (DR8) and our best fit model of the galaxy spectrum (red) using XSL population templates.  SDSS has the best signal-to-noise of the large-scale surveys available for our purposes, so this comparision is the most stringent test of the accuracy of our modeling.  The gap between 5700 \AA\ and 6100 in the best fit spectrum is due to the dichroic issue in X-shooter (see \cite{Vernet11}). We note that because of the parameter coverage (i.e., $T_{eff}$, log$(g)$, and $[Fe/H]$) of the first release of XSL, the fitting in the red part of the spectrum is not quite right in some cases, but this does not influence the wavelength ranges used for AGN identification.  We achieved a good fit of the absorption features and can, therefore, subtract the absorption from the AGN emission lines (i.e., \Ha, \Hb,  \oiiilam, \niilam). After the best fit is subtracted from the original spectrum, the emission line features are cleaned and identified. 
We show the residual spectra of the wavelength regions of interest in panels (b) and (c), with the emission lines labeled. To identify the emission line features quantitatively, we make use of the residuals in regions which are free from either bad pixels or emission lines. A mean rms is calculated and its flux in each emission line region is used to quantify the significance of the emission features. We define a significant emission line as one with a flux larger than 1.64 times its rms (90\% C.L.). We define an `emission line galaxy' as one with two or more significant emission lines. 

\begin{wrapfigure}[21]{r}{0.5\textwidth}
\vspace*{-0.9cm}
\includegraphics[width=0.45\textwidth, angle=90]{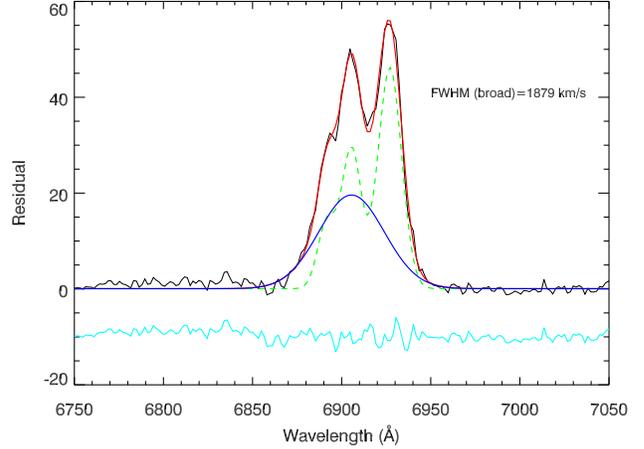}
\vspace{-0.9cm}
\caption{\small Emission line profile fitting around the $\rm{H_{\alpha}}- \rm{[NII]}$ wavelength region for
a Type I AGN candidate.
The red line is a four-component Gaussian profile,  while the cyan line is the residual between data and
the best fit, shifted for showing purpose. The blue line shows the broad compnent of $\rm{H_{\alpha}}$, 
and the green dashed line
shows the narrow component of the triple Gaussian, i.e., narrow $\rm{[NII]6548}$, $\rm{H_{\alpha}}$, 
 and $\rm{[NII]6584}$, respectively. }
\label{gaussianfit}
\end{wrapfigure}

We estimate the flux and full-width at half maximum (FWHM) for each emission line from the residual spectrum by fitting Gaussian profiles. When the lines are blended together, e.g., $\rm{H_{\alpha}}-\rm{[NII] 6548 / 6584}$ in panel (c), we fit multiple (two to four) Gaussian components to account for all the broad and narrow lines. We show an example of four Gaussian fitting around the blended \nii-\Ha\ line regions in Figure~\ref{gaussianfit}. If the flux of the broad component of  \Ha\ is less than the integrated rms in the same wavelength region, we use  three Gaussians to fit the blended narrow lines, as shown. Type I AGN are identified by FWHMs $\ge 1000$ km/s of the broad component of  \Ha. Therefore the galaxy shown in Figure~\ref{gaussianfit} is very likely to be a Type I AGN. We note that it is possible for a weak broad component of \Ha\ to be unidentified in low signal-to-noise cases.

\begin{wrapfigure}[17]{r}{0.5\textwidth}
\vspace*{-1.5cm}
\includegraphics[width=0.45\textwidth, angle=90]{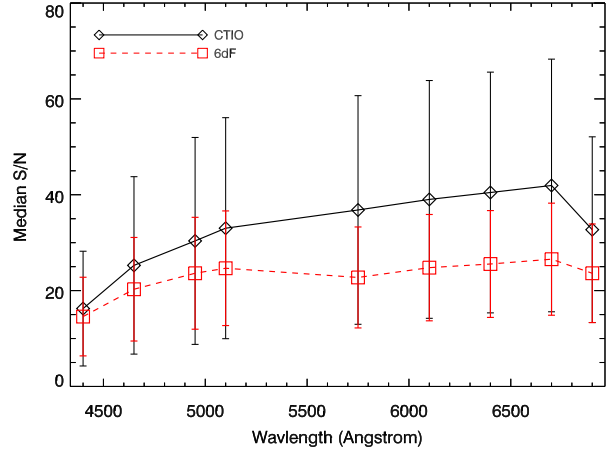}
\vspace*{-0.5cm}
\caption{Signal-to-noise ratios for CTIO and 6dF spectra at different wavelengths. While the average signal-to-noise ratio is higher for CTIO, the $1-\sigma$ error bars (1 standard deviation) overlap almost completely for the two samples, indicating that they are of similar quality.}
\label{SNRcomp}
\vspace*{-0.5cm}
\end{wrapfigure}

Since the signal-to-noise (S/N) is different for spectra taken with different instruments, the fraction of galaxies identified as emission line galaxies will be different for the different samples which make up 2MRS. Fortunately, the southern sky is mostly covered ($\sim$82\%) by CTIO and 6dF data and the spectra have comparable signal-to-noise over the full wavelength range as shown in Figure~\ref{SNRcomp}.  Adjusting for the differences in signal-to-noise of the more heterogeneous spectra available for the northern hemisphere will be the topic of a future paper.

When a broad line is not present, we use the narrow lines to identify AGN. The (relative) flux for each of the four lines is calculated from the Gaussian fits, the \oiiilam/\Hb\ and \niilam/\Ha\ ratios are calculated, and the BPT test is applied. To check our method for obtaining the emission line fluxes, we compare the flux we obtain for each of the four lines of interest for the SDSS sample to the fluxes determined by the SDSS team (DR8).  Space does not permit including plots here, but the fluxes are tightly correlated for emission line galaxies. The scatter in the relation is due to the fact that SDSS uses theoretical modeling for the galaxy subtraction while we use the XSL.   On average, we measure a higher flux for the emission lines than SDSS, indicating that our method is more sensitive to emission lines. 






\section{The Southern AGN Catalog}
\begin{figure}[htb]
\vspace*{-0.5cm}
\includegraphics[width=.475\textwidth, angle=90]{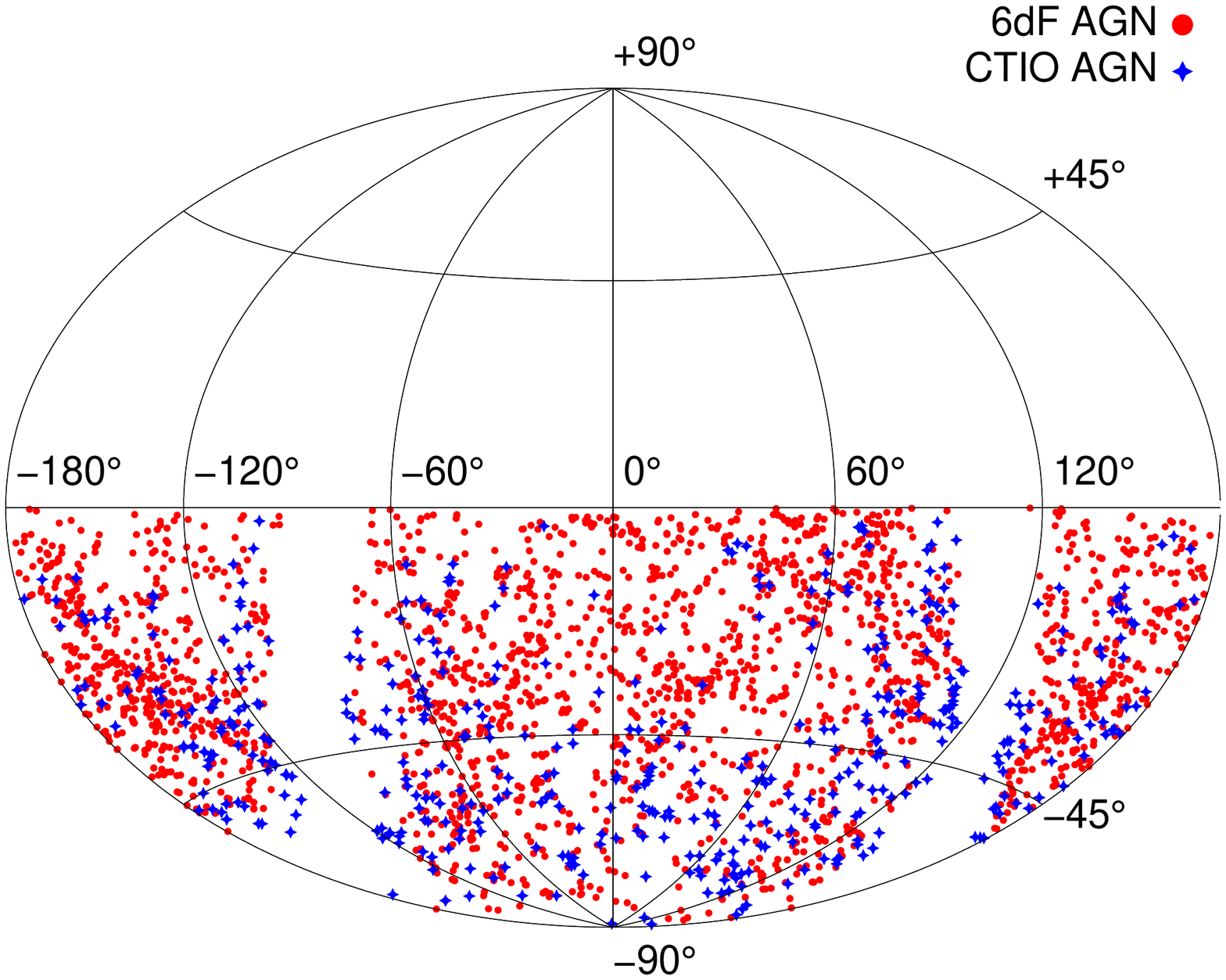}
\includegraphics[width=.475\textwidth, angle=0]{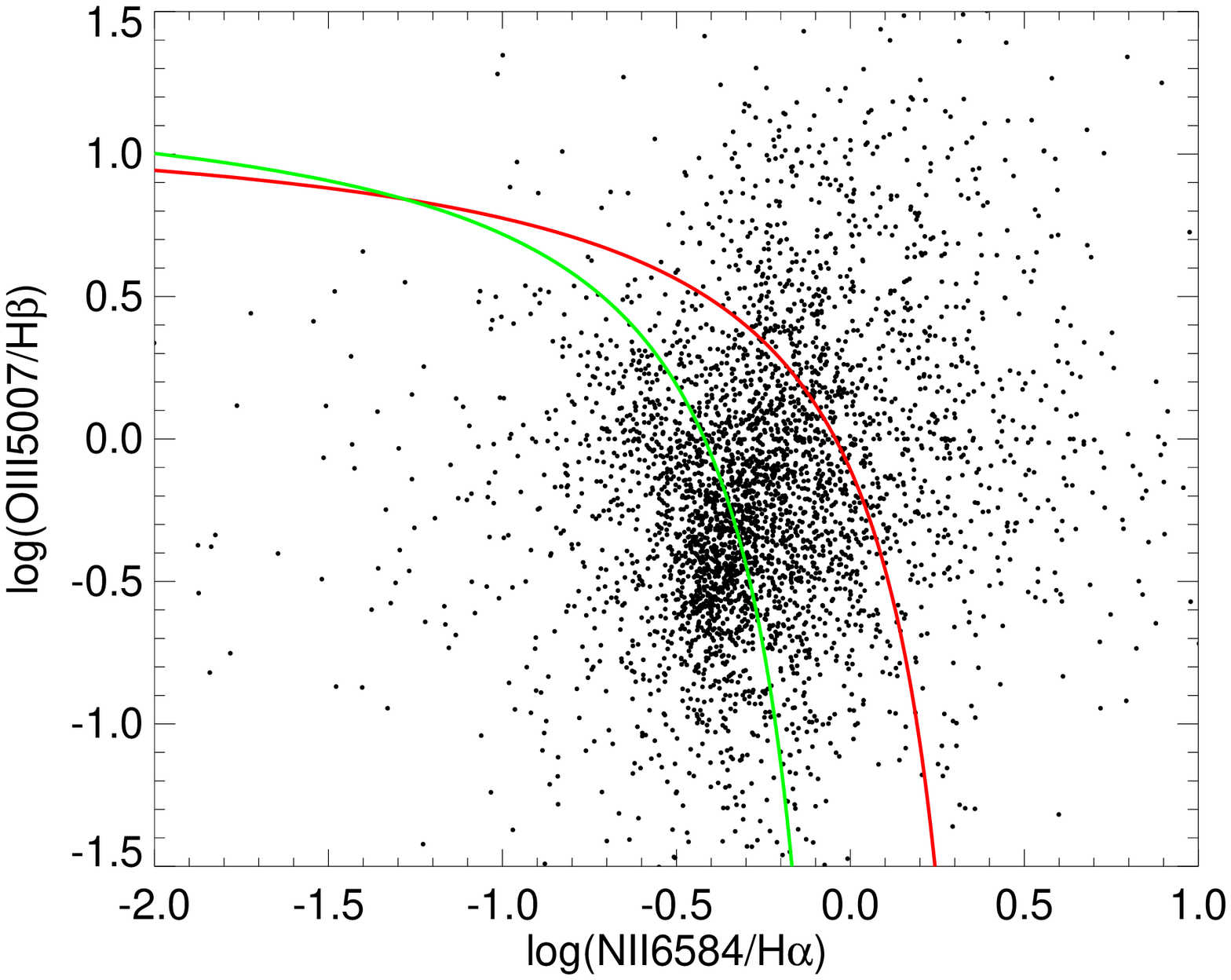}
\vspace*{-0.5cm}
\caption{Left panel: The sky distribution of the AGN (broad line AGN and the narrow line AGN satisfying the Kauffmann \cite{Kauffmann03} criteria) in our catalog. Right panel: The BPT diagram for the galaxies with at least two emission lines identified at 90\% C.L. The green and red lines denote the Kauffmann \cite{Kauffmann03} and Kewley \cite{Kewley01} criteria for narrow line AGN, respectively.}
\vspace*{-0.5cm}
\label{AGNskyplot}
\end{figure}

The sky distribution of our Southern AGN Catalog is shown in the left panel of Figure~\ref{AGNskyplot} and its BPT diagram is shown in the right panel.  The catalog consists of 569 broad line AGN, 832 (2026) narrow line AGN based on the Kewley \cite{Kewley01} (Kauffmann \cite{Kauffmann03}) criteria. The VCV catalog has only 814 AGN candidates (dec $ < 0^\circ$, $z < 0.0407$, $|b| > 10^\circ$), with a considerable fraction of them being non-AGN emission line galaxies \cite{zfg09,tzf12}, so our catalog contains many more robustly identified AGN as well as being uniformly selected. 

\begin{figure}[thb]
\includegraphics[width=0.35\textwidth, angle=90]{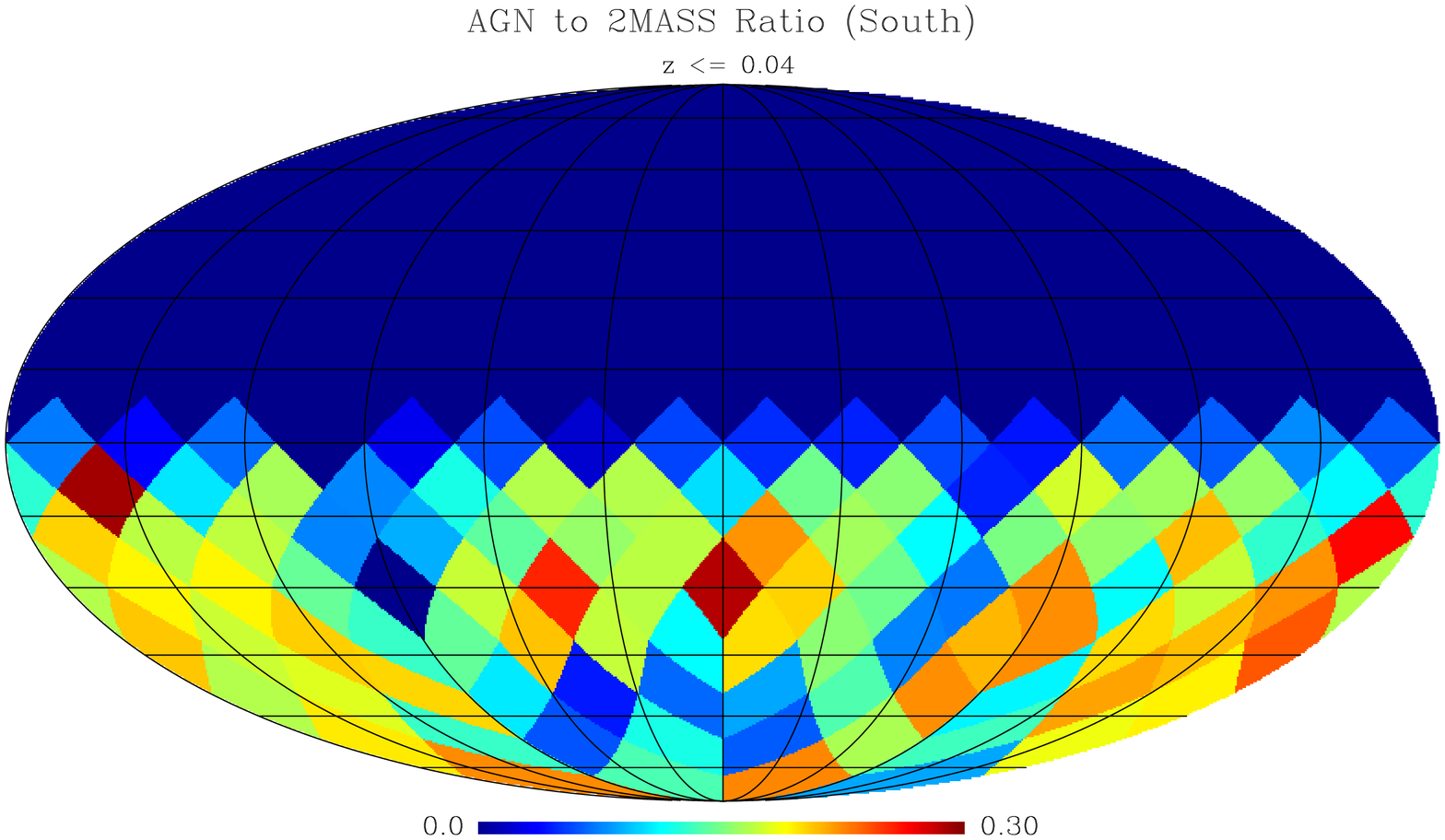}
\includegraphics[width=0.45\textwidth, angle=0]{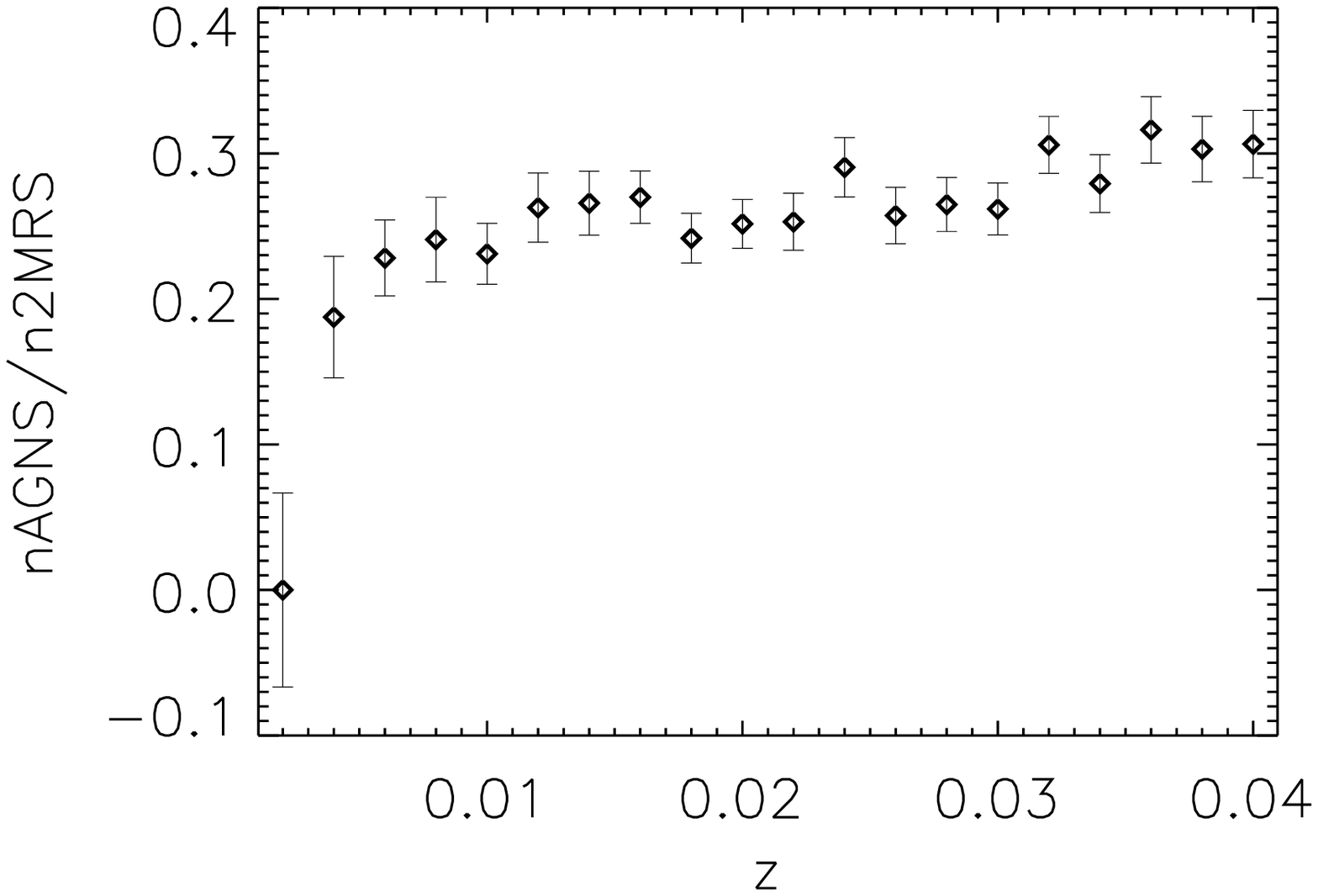}
\caption{AGN ratios across the sky (left panel, the sky has been divided into 192 equal area regions and the color indicates the AGN fraction), and with respect to redshift (right panel). The AGN ratios are roughly uniform, as expected assuming that AGN trace large scale structure.}
\label{AGNratio}
\vspace*{-1.0cm}
\end{figure}

We can assess the uniformity and completeness of this AGN sample by comparing the distribution of the AGN to the distribution of galaxies in 2MRS; because AGN trace large scale structure in general, the AGN-to-galaxy ratio should be uniform across the sky and in redshift. The analogous study of the VCV catalog \cite{zfb10} showed that the AGN fraction was highly non-uniform both across the sky and in redshift, with clumpy patterns in just a few regions and a high AGN fraction near-by ($z < 0.003$) and quickly dropping off at higher redshifts. Figure~\ref{AGNratio} shows the AGN-to-galaxy ratios for our catalog, across the sky (left panel, the sky has been divided into 192 equal area regions and the color indicates the AGN fraction.) and in redshift (right panel). In these plots, we have combined the broad line AGN with the narrow line AGN which satisfy the Kauffmann \cite{Kauffmann03} criteria to determine the AGN fraction relative to all the galaxies in 2MRS.  

The full catalog can be found at \newline
{\tt \footnotesize http://nyuad.nyu.edu/en/academics/faculty/ingyin-zaw/agn-catalog.html}. Our cuts were chosen to have a catalog that is as complete as possible while still being confident that we are identifying emission line galaxies.   In the online catalog we report the name, RA, Dec,  type of AGN, the fluxes and errors of the four lines, FWHM of the \Ha\ line, and the signal-to-noise ratio of the four lines for each of the AGN.   For a catalog with higher purity, the user can apply more stringent criteria, e.g. using only narrow line AGN satisfying \cite{Kewley01} criteria and requiring higher signal-to-noise ratios on the emission lines.  

\section{Summary}
We have constructed a uniformly selected catalog of nearby ($z \leq 0.0407$) optical AGN in the southern hemisphere outside the Galactic plane (dec $< 0^\circ$, $|b| > 10^\circ$), based on the 2MRS parent galaxy catalog. Using spectra taken by the 6dF Galaxy Survey and by the 2MRS team using the CTIO telescope, which are of similar quality, we identified 351 broad line AGN, 1914 narrow line AGN based on the Kewley \cite{Kewley01} criteria, and 4238 narrow line AGN based on the Kauffmann \cite{Kauffmann03} criteria. These AGN are distributed roughly uniformly across the sky and in redshift relative to the parent sample of 2MRS galaxies, as expected to the extent that AGN trace large scale structure. In addition to names, sky coordinates, and velocities of the identified AGN, we provide the line fluxes, flux errors, and signal-to-noise ratios, so that the user can apply their own selection criteria. This is currently the most complete and uniform optical AGN catalog available for the southern sky.  The northern sample has been processed and efforts are being made to assess and improve its uniformity and completeness. The website will be updated as progress is made.


\section*{Acknowledgments}
We thank Lucas Macri and Jessica Mink for sharing and explaining the FAST and CTIO spectra. We are grateful to Heath Jones, Philip Lah, and Matthew Colless for providing us with and helping us understand the 6dF spectra. We thank Lincoln Greenhill, Avanti Tilak, and Yanfei Zhang for useful discussions and assistance. The research of GRF was supported in part by the U.S. National Science Foundation (NSF), Grant PHY-1212538, and the James Simons Foundation. 


\providecommand{\href}[2]{#2}\begingroup\raggedright\endgroup

\end{document}